\documentstyle[11pt,epsf]{article}
\makeatletter
\def\revtex@ver{1.6}
\def\revtex@date{12 Aug 93}
\def\revtex@org{PASP}
\def\revtex@jnl{}
\def\revtex@genre{conference proceedings}
\def\revtex@pageid{\xdef\@thefnmark{\null}
\@footnotetext{This \revtex@genre\space was prepared with the
\revtex@org\space \revtex@jnl\space Rev\TeX\ macros v\revtex@ver.}}
\ifnum\@ptsize<1
\fi
\def\ps@paspcstitle{\let\@mkboth\@gobbletwo
\def\@oddhead{\null{\footnotesize\it\@slug}\hfil}
\def\@oddfoot{\rm\hfil\thepage\hfil}
\let\@evenhead\@oddhead\let\@evenfoot\@oddfoot
}
\def\ps@myheadings{\let\@mkboth\@gobbletwo
\def\@oddhead{\hbox{}\hfil\sl\rightmark\hskip 1in\rm\thepage}%
\def\@oddfoot{}%
\def\@evenhead{\rm\thepage\hskip 1in\sl\leftmark\hfil\hbox{}}%
\def\@evenfoot{}\def\sectionmark##1{}\def\subsectionmark##1{}}
\def\@leftmark#1#2{\sec@upcase{#1}}
\def\@rightmark#1#2{\sec@upcase{#2}}
\def\@singleleading{0.9}
\def\@doubleleading{1.6}
\def\baselinestretch{\@singleleading}
\def\tightenlines{\def\baselinestretch{\@singleleading}}
\def\loosenlines{\def\baselinestretch{\@doubleleading}}
\clubpenalty\@M
\widowpenalty\@M
\def\@journalname{ASP Conference Series}
\def\cpr@holder{Astronomical Society of the Pacific}
\def\@jourvol{10000}
\def\cpr@year{1994}
\def\vol@title{Astronomical Data Analysis Software and Systems III}
\def\vol@author{R.\ J.\ Hanisch, D.\ R.\ Crabtree, and J.\ Barnes, eds.}
\let\journalid=\@gobbletwo
\let\articleid=\@gobbletwo
\let\received=\@gobble
\let\accepted=\@gobble
\def\@slug{{\tabcolsep\z@\begin{tabular}[t]{l}\vol@title\\
\@journalname, Vol.\ \@jourvol, \cpr@year\\
\vol@author
\end{tabular}}
}
\def\paspconf@frontindent{.45in}
\def\title#1{\vspace*{1.0\baselineskip}
\@tempdima\textwidth \advance\@tempdima by-\paspconf@frontindent
\hfill
\parbox{\@tempdima}
	{\pretolerance=10000\raggedright\large\bf\sec@upcase{#1}}\par
\vspace*{1\baselineskip}\thispagestyle{title}}
\def\author#1{\vspace*{1\baselineskip}
\@tempdima\textwidth \advance\@tempdima by-\paspconf@frontindent
\hfill
\parbox{\@tempdima}
{\pretolerance=10000\raggedright{#1}}\par}
\def\affil#1{\vspace*{.5\baselineskip}
\@tempdima\textwidth \advance\@tempdima by-\paspconf@frontindent
\hfill
\parbox{\@tempdima}
{\pretolerance=10000\raggedright{\it #1}}\par}

\def\abstract{\vspace*{1.3\baselineskip}\bgroup\leftskip\paspconf@frontindent
\noindent{\bf\sec@upcase{Abstract.}}\hskip 1em}
\def\endabstract{\par\egroup\vspace*{1.4\baselineskip}}
\skip\footins 4ex plus 1ex minus .5ex
\long\def\@makefntext#1{\noindent\hbox to\z@{\hss$^{\@thefnmark}$}#1}

\def\tablenotetext#1#2{
\@temptokena={\vspace{.5ex}{\noindent\llap{$^{#1}$}#2}\par}
\@temptokenb=\expandafter{\tblnote@list}
\xdef\tblnote@list{\the\@temptokenb\the\@temptokena}}
\def\spewtablenotes{
\ifx\tblnote@list\@empty
\else
\let\@temptokena=\tblnote@list
\gdef\tblnote@list{\@empty}
\vspace{4.5ex}
\footnoterule
\vspace{.5ex}
{\footnotesize\@temptokena}
\fi}
\newtoks\@temptokenb
\def\tblnote@list{}
\def\endtable{\spewtablenotes\end@float}
\@namedef{endtable*}{\spewtablenotes\end@dblfloat}
\let\tableline=\hline
\def\thefigure{\@arabic\c@figure}
\def\fnum@figure{Figure \thefigure.}
\def\thetable{\@arabic\c@table}
\def\fnum@table{Table \thetable.}
\long\def\@makecaption#1#2{
\vskip 10pt
\setbox\@tempboxa\hbox{#1\hskip 1.5em #2}
\let\@tempdima=\hsize \advance\@tempdima by -2em
\ifdim \wd\@tempboxa >\@tempdima
	{\leftskip 2em
	#1\hskip 1.5em #2\par}
\else
	\hbox to\hsize{\hskip 2em\box\@tempboxa\hfil}
\fi}
\def\fps@figure{tbp}
\def\fps@table{htbp}
\let\keywords=\@gobble
\let\subjectheadings=\@gobble
\def\upper{\def\sec@upcase##1{\uppercase{##1}}}
\def\sec@upcase#1{\relax#1}
\def\section{\@startsection {section}{1}{\z@}{-4.2ex plus -1ex minus
-.2ex}{2.2ex plus .2ex}{\normalsize\bf}}
\def\subsection{\@startsection{subsection}{2}{\z@}{-2.2ex plus -1ex minus
-.2ex}{1.1ex plus .2ex}{\normalsize\bf}}

\def\subsubsection{\@startsection{subsubsection}{3}{\z@}{-2.2ex plus
-1ex minus -.2ex}{-1.2em}{\normalsize\it}}
\def\thesection{\@arabic\c@section.}
\def\thesubsection{\thesection\@arabic\c@subsection.}
\def\thesubsubsection{\thesubsection\@arabic\c@subsubsection.}
\def\@sect#1#2#3#4#5#6[#7]#8{\ifnum #2>\c@secnumdepth
\def\@svsec{}\else
\refstepcounter{#1}\edef\@svsec{\csname the#1\endcsname\hskip 1em }\fi
\@tempskipa #5\relax
\ifdim \@tempskipa>\z@
\begingroup #6\relax
\@hangfrom{\hskip #3\relax\@svsec}{\interlinepenalty \@M \sec@upcase{#8}\par}%
\endgroup
\csname #1mark\endcsname{#7}\addcontentsline
{toc}{#1}{\ifnum #2>\c@secnumdepth \else
\protect\numberline{\csname the#1\endcsname}\fi
#7}\else
\def\@svsechd{#6\hskip #3\@svsec #8\csname #1mark\endcsname
{#7}\addcontentsline
{toc}{#1}{\ifnum #2>\c@secnumdepth \else
\protect\numberline{\csname the#1\endcsname}\fi
#7}}\fi
\@xsect{#5}}
\def\@ssect#1#2#3#4#5{\@tempskipa #3\relax
\ifdim \@tempskipa>\z@
\begingroup #4\@hangfrom{\hskip #1}{\interlinepenalty \@M \sec@upcase{#5}\par}
\endgroup
\else \def\@svsechd{#4\hskip #1\relax #5}\fi
\@xsect{#3}}
\def\acknowledgments{\@startsection{paragraph}{4}{1em}
{1ex plus .5ex minus .5ex}{-1em}{\bf}{\sec@upcase{Acknowledgments.}}}

\def\qanda@heading{Discussion}
\newif\if@firstquestion \@firstquestiontrue
\newenvironment{question}[1]{\if@firstquestion
\section*{\qanda@heading}\global\@firstquestionfalse\fi
\par\vskip 1ex
\noindent{\it#1\/}:}{\par}
\newenvironment{answer}[1]{\par\vskip 1ex
\noindent{\it#1\/}:}{\par}
\def\mathwithsecnums{
\@newctr{equation}[section]
\def\theequation{\hbox{\normalsize\arabic{section}-\arabic{equation}}}}
\def\references{\section*{References}
\bgroup\parindent=0pt\parskip=.5ex
\def\refpar{\par\hangindent=3em\hangafter=1}}
\def\endreferences{\refpar\egroup}

\def\@biblabel#1{\relax}
\def\@cite#1#2{#1\if@tempswa , #2\fi}
\def\reference{\relax\refpar}
\def\@citex[#1]#2{\if@filesw\immediate\write\@auxout{\string\citation{#2}}\fi
\def\@citea{}\@cite{\@for\@citeb:=#2\do
{\@citea\def\@citea{,\penalty\@m\ }\@ifundefined
{b@\@citeb}{\@warning
{Citation `\@citeb' on page \thepage \space undefined}}%
{\csname b@\@citeb\endcsname}}}{#1}}
\let\jnl@style=\rm
\def\ref@jnl#1{{\jnl@style#1\/}}
\def\aj{\ref@jnl{AJ}}
\def\araa{\ref@jnl{ARA\&A}}
\def\apj{\ref@jnl{ApJ}}
\def\apjl{\ref@jnl{ApJ}}
\def\apjs{\ref@jnl{ApJS}}
\def\ao{\ref@jnl{Appl.Optics}}
\def\apss{\ref@jnl{Ap\&SS}}
\def\aap{\ref@jnl{A\&A}}
\def\aapr{\ref@jnl{A\&A~Rev.}}
\def\aaps{\ref@jnl{A\&AS}}
\def\azh{\ref@jnl{AZh}}
\def\baas{\ref@jnl{BAAS}}
\def\jrasc{\ref@jnl{JRASC}}
\def\memras{\ref@jnl{MmRAS}}
\def\mnras{\ref@jnl{MNRAS}}
\def\pra{\ref@jnl{Phys.Rev.A}}
\def\prb{\ref@jnl{Phys.Rev.B}}
\def\prc{\ref@jnl{Phys.Rev.C}}
\def\prd{\ref@jnl{Phys.Rev.D}}
\def\prl{\ref@jnl{Phys.Rev.Lett}}
\def\pasp{\ref@jnl{PASP}}
\def\pasj{\ref@jnl{PASJ}}
\def\qjras{\ref@jnl{QJRAS}}
\def\skytel{\ref@jnl{S\&T}}
\def\solphys{\ref@jnl{Solar~Phys.}}
\def\sovast{\ref@jnl{Soviet~Ast.}}
\def\ssr{\ref@jnl{Space~Sci.Rev.}}
\def\zap{\ref@jnl{ZAp}}

\def\la{\mathrel{\hbox{\rlap{\hbox{\lower4pt\hbox{$\sim$}}}\hbox{$<$}}}}
\def\ga{\mathrel{\hbox{\rlap{\hbox{\lower4pt\hbox{$\sim$}}}\hbox{$>$}}}}

\newcount\lecurrentfam
\def\LaTeX{\lecurrentfam=\the\fam \leavevmode L\raise.42ex
\hbox{$\fam\lecurrentfam\scriptstyle\kern-.3em A$}\kern-.15em\TeX}
\def\plotone#1{\centering \leavevmode
\epsfxsize=\textwidth \epsfbox{#1}}

\def\plotfiddle#1#2#3#4#5#6#7{\centering \leavevmode
\vbox to#2{\rule{0pt}{#2}}
\includegraphics{#1}}
\newif\if@finalstyle \@finalstylefalse
\if@finalstyle
\ps@myheadings
\let\ps@title=\ps@paspcstitle
\else
\ps@plain
\let\ps@title=\ps@plain
\fi
\ds@twoside
\makeatother
\begin{document}
\thispagestyle
\markright{\tiny\noindent To appear in {\bf Barred Galaxies}, IAU Coll.~157,
R.~Buta, B.G.~Elmegreen \& D.A.~Crocker (eds.), ASP Series, (1996)}
%
\title{3D Barred Model of the Milky Way Including Gas}
\author{Roger Fux and Daniel Friedli}
\affil{Geneva Observatory, CH-1290 Sauverny, Switzerland}
%
\begin{abstract}
We present a 3D N-body simulation of the Milky Way including $4\cdot 10^5$
star- and dark-like particles and $2\cdot 10^4$ gas particles, initially
distributed according to an axisymmetric, observationally constrained, mass
model. The whole system is self-gravitating and the gas hydrodynamics is
solved using the SPH method. The simulation leads to the spontaneous formation
of a central bar that strongly affects the gas dynamics. We compute ($l-V$)
diagrams for both the gaseous and the stellar particles as a function of the
angle of the bar with respect to the observer and compare the results for the
gas with HI and CO observations.
\end{abstract}
%
\vspace{-.05cm}\section{Introduction}
%
Since the first direct evidences for the Galactic bar (Blitz \& Spergel 1991;
Weinberg 1992), gas and stellar dynamics in the Milky Way have experienced a
clear renewal of interest. Several new photometric and kinematic data have
already been or are on the way to be used to constrain analytical triaxial
bulge models (e.g. Stanek, this volume). By performing numerical simulations,
the problem can be tackled differently: time sampled snapshots of the
numerical galaxy can be compared to the observations varying various model
parameters (see e.g. Wada et al. 1994). Invaluable informations can then be
inferred concerning quantities hard to observe, like the bar pattern speed.
Nevertheless, so far none of the dynamical models have included in a
completely self-consistent way three dimensional gaseous, stellar, and dark
components. This is however a prerequisite to highlight various global
asymmetries or instabilities, or if the evolution of the model needs to be
followed for Gyrs since secular evolution processes are especially active in
barred galaxies (see e.g. Martinet 1995).
%
\vspace{-.05cm}\section{Initial Conditions}
%
Our initial axisymmetric 3D mass distribution includes four components (see
Table~1): an exponential stellar disk of constant thickness, a composite
power-law stellar nucleus-spheroid (NS), a dark halo (DH) to ensure an almost
flat rotation curve out to about 30~kpc, and a dissipative gas component
consisting~of a kind of smoothly truncated Mestel disk with a central core and
an outward linearly increasing scale height. The gas mass inside the solar
circle is $4.1\cdot 10^9$~M$_{\odot}$. The parameters of this mass model have
been adjusted to several observational constraints, like the rotation curve
and local mass densities, assuming $R_{\circ}\!=\!8$~kpc. To each of the
stellar and dark components is assigned an individual isotropic Maxwellian
velocity distribution such that the velocity dispersion and the mean azimuthal
velocity (no streaming motion in other directions) satisfy the Jeans
equations. Initially, the gas particles have circular velocities and a uniform
sound speed of 10~km/s, providing only a rough hydrostatic equilibrium. The
non-inclusion of star formation presently is the strongest limitation of the
model.
\begin{table}[t]
\caption{Initial mass model of the simulation. The oblate NS and~DH have a
common axis ratio $e\!=\!0.5$.}
\vspace{-.1cm}
\begin{center}\small
\begin{tabular}{llll}
\tableline
 \vspace{-.3cm} & & & \\
Components\hspace*{-.2cm} & \# particles & Space density & Parameters \\
 \vspace{-.3cm} & & & \\
 \tableline
 \vspace{-.2cm} & & & \\
Disk & $150,000$
   & $\propto \exp{(-R/h_R)}\cdot\hbox{sech}\hspace{.1cm}(z/h_z)$
   & $h_R = 2.5$ kpc \\
   & & & $h_z = 250$ pc \\
   & & & $M = 4.4 \cdot 10^{10}$ M$_{\odot}$\\
NS & \phantom{0}$50,000$
   & ${\displaystyle \propto \frac{m^p}{1+m^{p-q}}}$,
   & $p = -1.8$,\hspace{.2cm} $q = -3.3$\\
 \vspace{-.2cm} & & & \\
   & & \hspace{.5cm} $m^2 = (R^2+z^2/e^2)/a^2$  & \\
 \vspace{-1.13cm} & & & \\
   & & & $a = 1$ kpc \\
   & & \vspace{.2cm} & $M = 3.0 \cdot 10^{10}$ M$_{\odot}$ \\
DH & $200,000$ & $\propto \exp{(-\mu)}$, & $b = 9.4$ kpc \\
   & & \hspace{.5cm} $\mu^2 = (R^2+z^2/e^2)/b^2$
   & $M = 3.1 \cdot 10^{11}$ M$_{\odot}$ \\
 \vspace{-.2cm} & & & \\
Gas & \phantom{0}$20,000$
   & $\propto \frac{1}{R\sqrt{R^2+h_g^2}}\exp{(-\frac{3}{2}\frac{R^2}{R_g^2}
    -\frac{1}{2}\frac{z^2}{s^2R^2})}$ & $s = 0.017$ \\
 \vspace{-.6cm} & & & \\
   & & & $h_g = \frac{1}{2}(h_R+b)$ \\
   & & & $R_g = 20$ kpc \\
   & & & $M = 1.13 \cdot 10^{10}$ M$_{\odot}$ \\
 \vspace{-.3cm} & & & \\
 \tableline
\end{tabular}
\end{center}
\vspace{-.35cm}
\end{table}
\begin{figure}
\plotfiddle{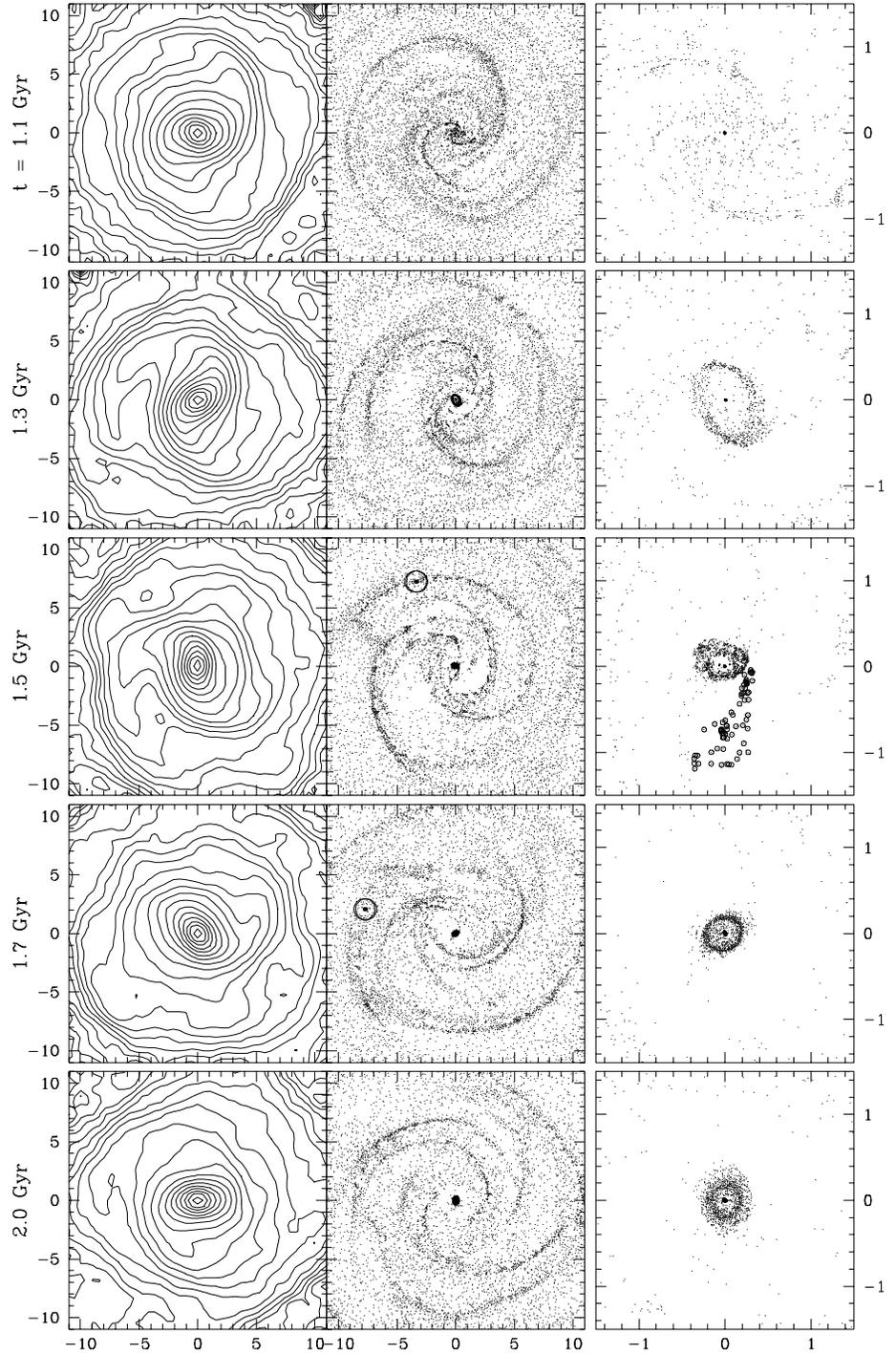}{18.3cm}{0}{102}{102}{-286}{-178.5}
\caption{\protect\small Face-on configuration of the visible components at
various times. Left: stellar (disk+NS) isodensity curves. Middle: gas
particles, same scale. The position of the observer in Figs.~2d and 3b is
indicated by a solid circle. Right: zoom on the central gas. The encircled
dots at 1.5~Gyr show the par- ticles responsible for the elongated strip in
Fig.~3b. The length unit is 1~kpc.}
\end{figure}
%
\vspace{-.35cm}\section{Time Evolution}
%
The system has been integrated for 2.6~Gyr using the PM method on
polar-cylindrical grid to compute the gravitational forces (Pfenniger \&
Friedli 1993) and the Lagrangian SPH technique (see e.g. Benz 1990) with
isothermal internal energy to compute the pressure and viscous forces acting
on the gas. The central radial resolution of the grid is 38~pc and the gas
smoothing length near the center is of the same order. The time integrator is
a synchronized leap-frog with an adaptative time-step $\Delta t \leq 0.1$ Myr
able to fully resolve shocks (variable $\Delta t$ in time but constant for all
particles). {\it The evolution is completely self-consistent and free from any
imposed symmetry}.
\par The simulation, illustrated in Fig.~1, leads to a growing central bar
after about 1~Gyr, while $m\!=\!1$ modes appear in the stellar disk. The bar
itself never really behaves like a bi-symmetric solid rotating body, but
instead undergoes many asymmetrical distortions, particularly strong during
its first rotations. Such perturbations also induce irregular gas flows (like
the one-side void inside 3~kpc occurring at $t\!=\!1.5$~Gyr) capable of
producing peculiar signatures in the gaseous ($l-V$) diagrams. The gas
particles near the center condense into a nuclear ring which progressively
shrinks and finally forms a small disk of about 300~pc radius. In this
process, the ring suffers several instabilities, like tilting, warping and
off-centering. There is also an increasing central gas core accreting gas from
the nuclear ring/disk. The final properties of the bar are a semi-major axis
$a \approx 3.5$~kpc, an axis ratio $b/a \approx 0.6-0.7$ and a pattern speed
$\Omega_p \approx 50$~km/s, which places the CR around 4~kpc, the ILR slightly
outside 1~kpc and the OLR close to 7~kpc. A comparison with symmetrized
simulations (i.e. imposed reflexion symmetries about the $z$-axis and the
$z\!=\!0$ plane) without gas (Fux et al. 1995) and with gas indicates that the
imposed symmetries generate more extended bars, whereas the presence of gas
makes them slightly rounder.
\begin{figure}[thbp]
\plotone{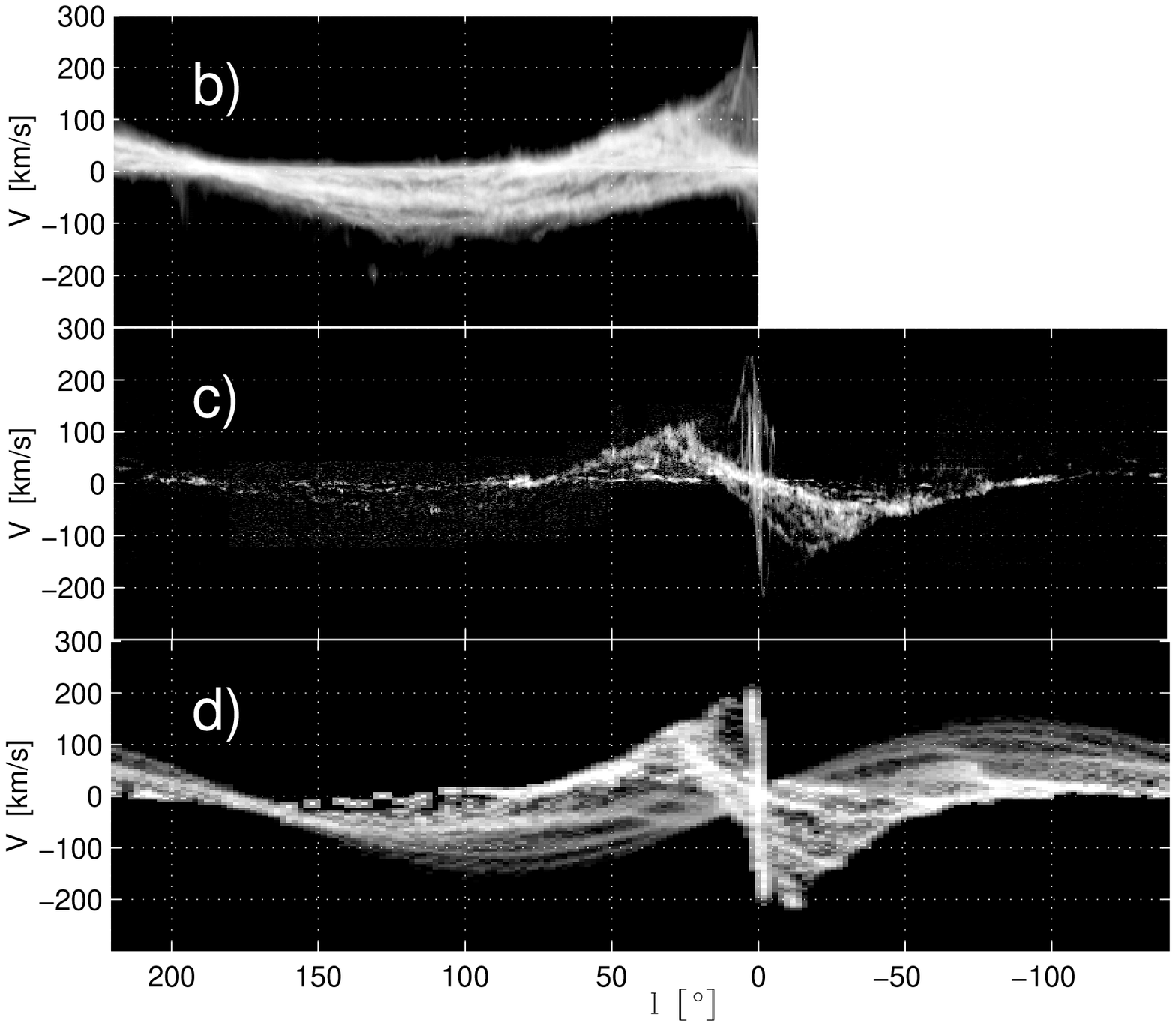}\vspace{-1cm}
\plotfiddle{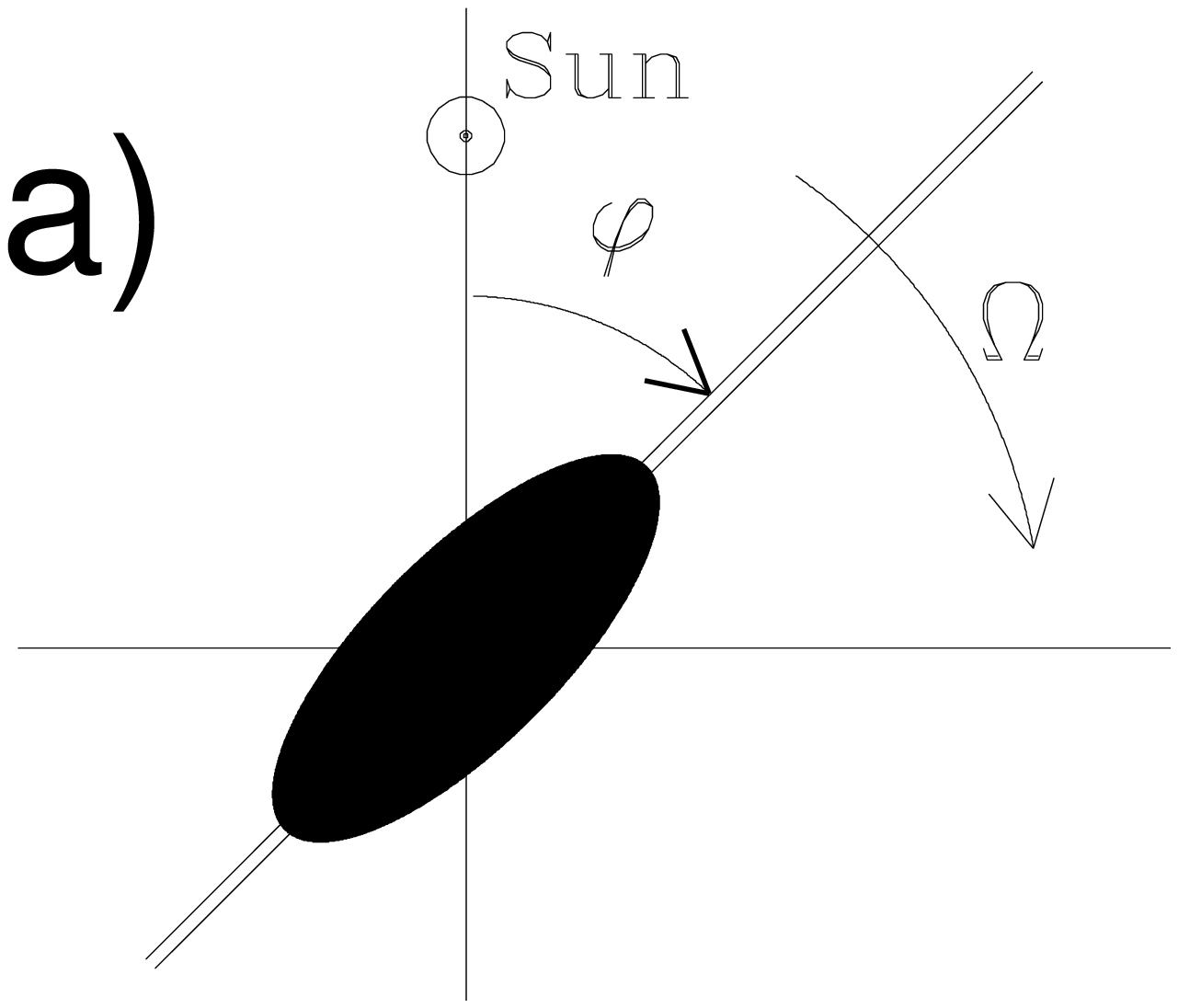}{0cm}{0}{30}{30}{13}{130}
\caption{a) Definition of the angle $\varphi$ of the bar relative to the
observer. $\varphi$ is positive in the direction of rotation. b) HI ($l-V$)
map for $0\leq~l~\leq~220^{\circ}$ (Hartman \& Burton 1995). c) CO ($l-V$) map
(Dame et al. 1987). d) Distance weighted ($l-V$) diagram from the simulation
at $t\!=\!1.7$~Gyr and for $\varphi \!\approx\!  30^{\circ}$. All ($l-V$)
plots are averaged over $|b| \!<\! 2.25^{\circ}$. Higher density regions are
whiter.}
\end{figure}
%
\vspace{-.05cm}\section{Gaseous ($l-V$) Diagrams}
%
For the moment, the longitude-velocity diagrams of the gas particles have only
been explored every 100 Myr, as a function of the relative angle $\varphi$ of
the bar with respect to the observer located at $R\!=\!R_{\circ}$ (see
Fig.~2a). It comes out that for a given angle $\varphi$, these diagrams are
time-dependent. Moreover, for each value of~$\varphi$ \hfill there \hfill is
\hfill a \hfill modulo \hfill $\pi$ \hfill choice \hfill in \hfill orienting
\hfill the \hfill bar, \hfill allowing \hfill for \hfill two \hfill possible
\newpage
\begin{figure}[thbp]
\plotone{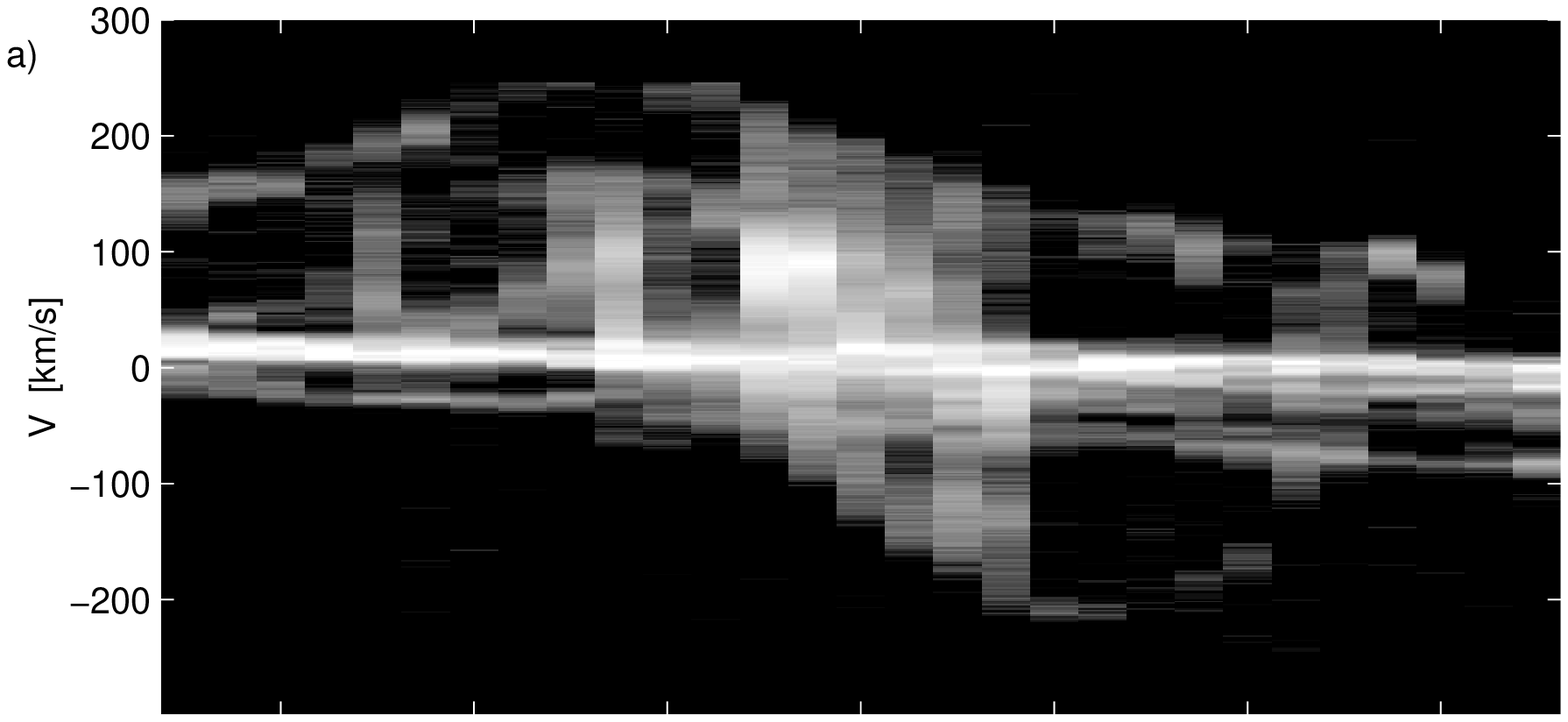}\\
\plotone{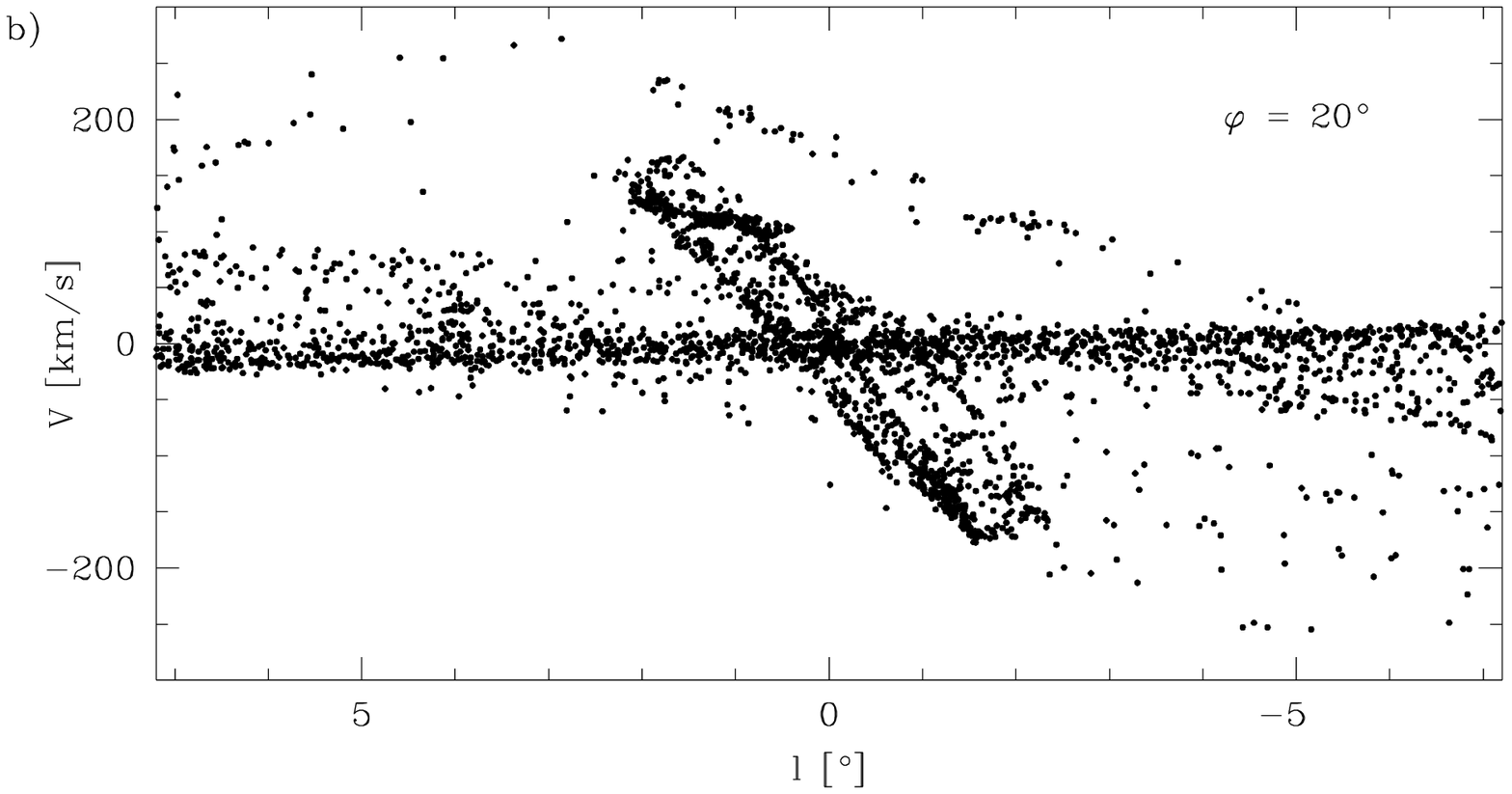}
\vspace{-.1cm}\caption{a) Details of the CO ($l-V$) map in the Galactic
central region. Note the positive high velocity emission strip reaching very
negative longitudes. b) One of the most resembling gaseous ($l-V$) plot
available from the simulation at $t\!=\!1.5$~Gyr and with
$\varphi \approx 20^{\circ}$, restricted to the gas particles inside
$|b|<2.25^{\circ}$.}
\end{figure}
\vspace{.15cm}
\hspace{-.9cm}
($l-V$) diagrams. These would look identical under reflexion symmetry about
the $z$-axis, but can appear very different in our symmetry-free case.
\par When computing ($l-V$) diagrams over a large longitude range in order to
confront them with observations, one should weight the contribution of each
particle by its inverse squared distance relative to the observer. Figure~2
shows such a diagram for a realistic position angle of the bar at a time where
similarities with the observed HI and CO maps clearly emerge. The gas in our
simulation can be considered as a mixture of atomic and molecular gas.
\par A well-known puzzling feature in the observed ($l-V$) data around the
Galactic center is the presence of gas in regions forbidden to pure circular
motion. Part of it can be explained by the signature of bar elongated $x_1$ or
anti-bar $x_2$ periodic orbits (Binney et al. 1991) or by more sophisticated
gas flow modeling (Mulder \& Liem 1986; Jenkins \& Binney 1994) in a fixed bar
potential. However, these approaches do not account for the intriguing gas
strip at $V \leq 100$~km/s extending in the negative longitude region down to
about $l\!=\!-5^{\circ}$ (see Fig.~3a). Our simulation displays at least once
such a peculiarity, as shown in Fig.~3b. The spatial positions of the gas
particles populating the ``forbidden'' strip in this figure are highlighted in
Fig.~1. A closer look to the simulation reveals that these particles were
previously strongly shocked near the bar end, then quickly fell toward the
center to finally overtake it with high positive radial velocities. Clearly,
the high degree of asymmetry displayed by the observations and achieved here,
for instance at $t\!=\!1.5$~Gyr, cannot occur in standard bi-symmetric bar
models.
%
\vspace{-.1cm}\section{Stellar ($l-V$) Diagrams}
%
Constraining the bar properties using stellar ($l-V$) diagrams will probably
prove to be at least as powerful as the methods based on gas dynamics. Since
observations in this field mainly rely on star counts, corresponding model
diagrams do not have to take into account a distance weighting factor as for
the gas. Fig.~4 gives a first example of what can be expected from such
diagrams. The signatures obviously depend on the bar orientation. A detailed
confrontation with observations will be published elsewhere.
\begin{figure}[bhtp]
\plotone{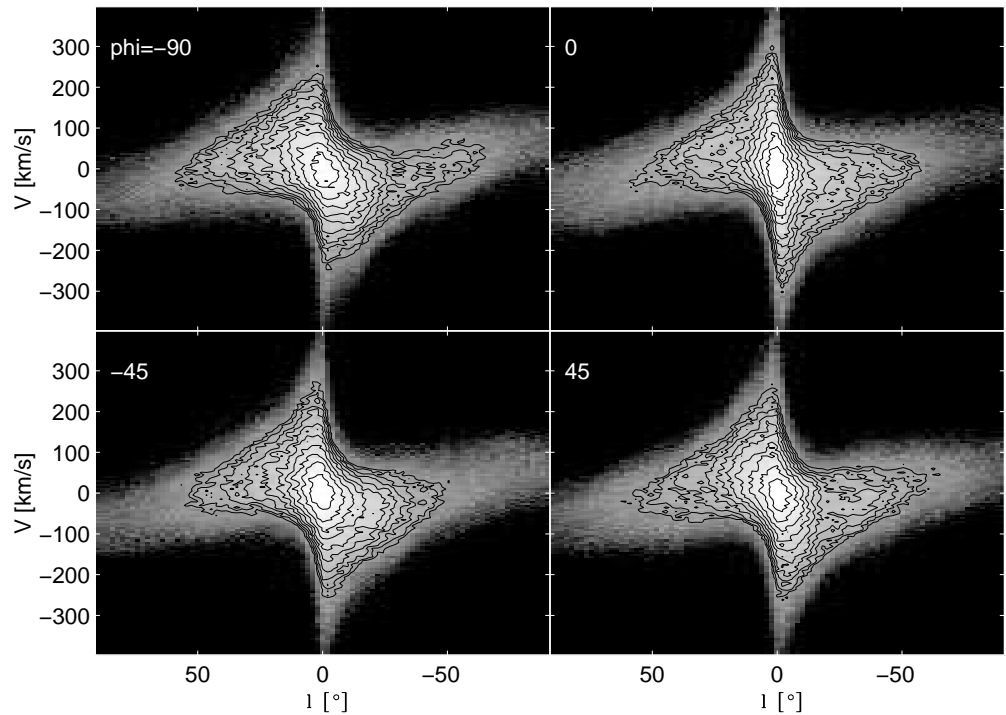}
\vspace{-.1cm}\caption{Distance unweighted ($l-V$) diagram of the stellar disk
as a function of $\varphi$ at $t\!=\!2.5$~Gyr, averaged over all $b$. Diagrams
for $\varphi$ and $\varphi + \pi$ have been added together to reduce noise.}
\end{figure}
%
\section{Conclusion}
%
A fully consistent numerical 3D barred model of the Milky Way in good
qualitative agreement with HI and CO kinematic observations at both small and
large longitudes has been presented. In particular, morphological and
kinematical asymmetries naturally occur during the simulation which are able
to reproduce the high velocity CO feature at negative longitudes in the
($l-V$) diagram.
%
\begin{question}{Christodoulou}
Your simulation run for just a few Gyrs. If you managed to take it to 15~Gyr,
which of the features would you expect to survive (bar, spirals, nuclear
ring)? How much mass is accumulated in the nuclear ring?
\end{question}
\begin{answer}{Fux}
Integration over a Hubble time would be unrealistic because the initial
conditions mimic the {\it present} state of the Milky Way and the absence
of star formation favor the growing of the central gas concentration, thus
shortening the life-time of the bar. At 1.5 Gyr, the ring mass is about
$6\cdot 10^8$ M$_{\odot}$.
\end{answer}
\begin{question}{Stanek}
Is your evolved disk similar to the real disk of the Galaxy (i.e. scale
height, velocity ellipsoid, etc.)?
\end{question}
\begin{answer}{Fux}
At $t=1.5$ Gyr, the scale height of the stellar disk is about 350~pc outside
corotation and its maximum l.o.s. velocity dispersion towards Baade's Window
about 130 km/s. Its velocity ellipsoid has become spontaneously anisotropic
with local velocity dispersions close to the observed ones for the Galactic
old~disk.
\end{answer}
\begin{question}{Teuben}
To what extent are the features like a tilted or offset nuclear disk due to
$\sqrt{N}$-noise, or robust. Is the tilt related to vertical instability
strip?
\end{question}
\begin{answer}{Fux}
They are robust with a density contrast far above the one resulting from
noise fluctuations. A vertical instability strip does exist, but we haven't
yet checked that the associated orbit families are consistent with the tilt
morphology.
\end{answer}
%

%
\end{document}